\documentclass[twocolumn,showpacs,preprintnumbers,amsmath,amssymb]{revtex4}
\usepackage[pdftex]{graphicx}
\usepackage{longtable}
\usepackage{dcolumn}% Align table columns on decimal point
\usepackage{bm}% bold math
\usepackage{color}
\usepackage{subeqnarray}
\definecolor{grey}{gray}{0.75}

\begin{document}
\title{Infrared study of the phonon modes in PrMnO$_3$ and CaMnO$_3$}
\author{R. Sopracase, G. Gruener, E. Olive and J.C. Soret}
\affiliation{Universit\'{e} Fran\c{c}ois Rabelais - LEMA UMR 6157 CNRS - CEA \\ Parc de
Grandmont - 37200 Tours - France}
\date{\today}

\begin{abstract}
The infrared (IR) reflectivity spectra of orthorhombic manganese perovskites PrMnO$_3$ and CaMnO$_3$ are studied in the frequency range of optical phonon modes at temperatures varying from 300 to 4 K. The IR phonon spectra of these two materials are analyzed by a fitting procedure based on a Lorentz model, and assigned to definite vibrational modes of $Pnma$ structures by comparison with the results of lattice dynamical calculations. The calculations have been performed in the framework of a shell model using short range Born-Mayer-Buckingham and long range Coulomb potentials, whose parameters have been optimized in order that the calculated Raman and IR active phonon frequencies, and lattice parameters match with their experimental values. We find a close correspondence between the values of the IR phonon frequencies of PrMnO$_3$ and CaMnO$_3$, which shows that the substitution of the Pr$^{3+}$ ions with Ca$^{2+}$ results in a reduction of the frequency of medium- and high-energy IR phonons, and an increase of the frequency of those of low-energy. Nevertheless, the experimentally obtained IR phonon amplitudes of the two materials appear to be unrelated. A comparative study of the vibrational patterns of these modes reveals that most of them correspond to complex atomic vibrations significantly different from PrMnO$_3$ to CaMnO$_3$ which cannot be assigned only to a given type of vibration (external, bending, or stretching modes). In particular, these results confirm that the structure of CaMnO$_3$ is quite far from the ideal (cubic) perovskite structure.
\end{abstract}

\maketitle

\section{Introduction}

Among the Mn oxides with perovskite structure \cite{Dagotto03} (the so-called manganites), Pr$_{1-x}$Ca$_x$MnO$_3$ is well known as the unique showing insulator behaviour over the whole range of doping ($x$) at zero magnetic field. This is usually assigned to a strong lattice distorsion, namely the tilting of MnO$_6$ octahedra (or an average Mn-O-Mn angle smaller than 180$^\circ$) induced by the small size of the Pr and Ca ions, and leading to the localization of the itinerant carriers in the $e_g$-orbitals. The phase diagram of Pr$_{1-x}$Ca$_x$MnO$_3$ exhibits a wide variety of ground states \cite{Jirak85, Yoshizawa95, Tomioka96, Martin99}. In particular for $0.3\le x\le0.75$, a $CE$-type charge ordered state --- where the Mn$^{3+}$ and Mn$^{4+}$ ions are alternately arranged within the $x$-$z$ planes (in the $Pnma$ coordinate axes orientation) with a concomitant ordering of the $e_g$-orbitals --- is stabilized below $T_{CO}=250$--200 K \cite{Jirak85,Zimmermann01} and then a $CE$-type antiferromagnetic spin arrangement takes place at $T_N=170$--180 K \cite{Jirak85}. An interesting point is that $T_N$ is smaller than $T_{CO}$. This suggests the existence of a strong electron-phonon interaction arising from the Jahn-Teller (JT) distorsions, while the exchange coupling between localized spins $S=\frac{3}{2}$ in the $t_{2g}$-orbital is lower. Note also that at low temperature when $0.3 \le x <0.5$, the $CE$-type spin ordering turns into a canted antiferromagnetic state ($T_{CA}<100$ K), which could consist of a mixture of ferromagnetic and antiferromagnetic clusters \cite{Jirak85}. External forces --- like magnetic fields \cite{Yoshizawa95, Tomioka96, Tomioka95, Lees95, Tokunaga98, Anane99}, electric fields\cite{Asamitsu97, Ponnambalam99, Parashar00, Stankiewicz00,Parashar04, Westhauser06}, hydrostatic pressure \cite{Hwang95, Moritomo97}, and photon exposure \cite{Kiryukhin97, Ogawa98}--- can destabilize the $CE$-type charge/orbital and spin ordered state into the ferromagnetic metallic phase that is absent in the zero magnetic field phase diagram of Pr$_{1-x}$Ca$_x$MnO$_3$. It is generally accepted that the competition between ferromagnetism and antiferromagnetism is an important key to the colossal magnetoresistance effect of the manganites. Such a phase competition results from the complex interplay of charge, orbital, spin, and lattice degrees of freedom. The latter is expected to be quite active in Pr$_{1-x}$Ca$_x$MnO$_3$ which is a strongly orthorhombically distorted manganites. 

Infrared (IR) spectroscopy is a very effective tool to study systems with strong electron-lattice coupling. Okimoto \textit{et al.} \cite{Okimoto99} measured optical spectra of a single crystal of Pr$_{1-x}$Ca$_x$MnO$_3$ ($x=0.4$) varying temperature and magnetic field. In particular, an increase of the number of optical phonons was observed at the charge/orbital transition, but their origin was not clearly established. Similarly phonon anomalies were observed at low temperature in the half-doped coumpound Pr$_{0.5}$Ca$_{0.5}$MnO$_3$ by Ta Phuoc \textit{et al.} \cite{TaPhuoc03}. Classifying the optical phonons of the end-member compounds PrMnO$_3$ and CaMnO$_3$ may be particularly useful for further studies on the evolution of these modes in Pr$_{1-x}$Ca$_x$MnO$_3$, and for identifying the additional phonon modes which should emerge in the charge/orbital ordered state.

The structural determination of PrMnO$_3$ was obtained by neutron diffraction \cite{Cherepanov, Jirak97a, Alonso00}, and by X-ray diffraction \cite{Cherepanov}. Below $\sim 900$ K, PrMnO$_3$ with the \textit{Pnma} orthorhombic structure presents a strong JT cooperative distorsion of MnO$_6$ octahedra related to the ordering of the $e_g$-orbitals \cite{Martin01}, and undergoes at $T_N \approx 95$ K an $A$-type antiferromagnetic spin ordering \cite{Jirak97b}. The structure of CaMnO$_3$ is also described with the $Pnma$ space group at temperatures ranging up to at least 800 K \cite{Zhou06}. However, available structural data \cite{Blasco00, Jorge} indicate that CaMnO$_3$ has an orthorhombically distorted perovskite structure much less pronounced than PrMnO$_3$. This is a consequence of the substitution of Mn$^{3+}$ ($t_g^3\:e_g^1$) with Mn$^{4+}$ ($t_g^3$) that actually induces two effects of different origins. One arises from the fact that Mn$^{4+}$ has no orbital degeneracy, and therefore does not produce any JT distorsion. The other is due to the mismatch between the ionic radii $r_R$ of the $R$-ions (Pr$^{3+}$ or Ca$^{2+}$) and the ionic radii $r_{Mn}$ of the Mn$^{3+}$ or Mn$^{4+}$ ions, which is measured by the tolerance factor $t=(r_{R}+r_O)/ \sqrt{2}(r_{Mn}+r_O)$ (where $t=1$ corresponds to an ideal perovskite structure, whereas $t<1$ indicates an average Mn-O-Mn angle that gets smaller than 180$^\circ$). Ca$^{2+}$ and Pr$^{3+}$ having similar ionic radius and those of Mn$^{4+}$ (0.53 \AA) and Mn$^{3+}$ (0.645 \AA) being significantly different, the tolerance factor of PrMnO$_3$ (0.94) turns out to be smaller than that of CaMnO$_3$ ($\simeq 1$), which should appear to have a structure very close to the ideal cubic perovskite. Recently, the Raman spectra of PrMnO$_3$ \cite{Martin01, Iliev06a, Laverdiere06a, Laverdiere06b, Martin03, Martin02} and CaMnO$_3$ \cite{Martin02, Liarokapis99, Granado01, Abrashev02, Iliev03} have been studied in detail. In particular, it appeared that the two dominant peaks in the Raman spectrum of PrMnO$_3$ are activated by stretching of the basal oxygen ions (O2) of the MnO$_6$ octahedra due to the JT effect, in striking contrast with the spectrum of CaMnO$_3$ where the two main peaks are activated by rotation of the octahedra around the $\lbrack101\rbrack$ cubic axis. Similarly, IR-active phonon modes in these materials are expected to be very sensitive to different types of distorsion (rotations of MnO$_6$ octahedra around the $\lbrack101\rbrack$ or $\lbrack010\rbrack$ cubic directions, JT distorsion, and shift of $R$ atoms)\cite{Abrashev02}. However, there has not been any systematic study of IR-active phonons in PrMnO$_3$, at least to our knowledge. Concerning CaMnO$_3$, reflectivity spectra of La$_{1-x}$Ca$_x$MnO$_3$ have been measured from 5 meV up to 30 eV for different values of $x$ including $x=1$, but only the electronic part of these spectra was discussed \cite{Jung98}. Furthermore, Fedorov \textit{et al.} \cite{Fedorov99} have presented IR absorption spectra of CaMnO$_3$, but the assignement of the phonon peaks was done by comparison with the IR absorption spectra of LaMnO$_3$ which is a JT compound.
   
In this paper, we present an analysis of IR-active phonons in PrMnO$_3$, and CaMnO$_3$. The experimentally observed phonon peaks are assigned to definite atomic vibrations by comparison with the results of lattice dynamical calculations (LDC) for both these materials.

 \section{Experimental details}
 
Very fine powders of high chemical homogeneity of PrMnO$_3$ and CaMnO$_3$ were prepared using a modified citrate process as described in Ref.\,\onlinecite{Douy01}. The polycrystalline sample of PrMnO$_3$ was obtained from these particles of PrMnO$_3$ pressed isostatically under 3 kbar in the form of a cylindrical slab of about 1 cm in diameter and 3 mm in thickness, and sintered at 1250$^\circ$C in air for 12 hours. The polycrystalline sample of CaMnO$_3$ was synthesized in the same way, but it was sintered at 1300$^\circ$C in air for 24 hours, and then annealed at 800$^\circ$C in oxygen current flow for 1 hour. Powder X-ray diffraction profiles recorded at room-temperature revealed that both samples had single-phase orthorhombic $Pnma$ structure with the lattice parameters $a=5.7852$ \AA,  $b=7.5939$ \AA, $c=5.4490$ \AA\ for PrMnO$_3$, and $a=5.2785$ \AA,  $b=7.4435$ \AA, $c=5.2585$ \AA\  for CaMnO$_3$. These data are in good agreement with the lattice parameters of stoichiometric samples reported previously \cite{Cherepanov, Jirak97a, Alonso00, Zhou06, Jorge, Blasco00,Poeppelmeier82}. Another indication of the quality of our samples was obtained from Raman measurements taken at room-temperature, which showed that both samples have Raman spectra very similar to those reported in the literature \cite{Martin01, Iliev06a, Laverdiere06a, Laverdiere06b, Martin03, Martin02, Liarokapis99, Granado01, Abrashev02, Iliev03}.

We have measured near normal incidence reflectivity spectra with a Bruker IFS 66v/S in the range 50 cm$^{-1}$- 17000 cm$^{-1}$ from 4 K to 300 K with a continuous flow cryostat. The samples were polished up to an optical grade of 0.25 $\mu$m.  After the measurement, the samples were coated with a silver film and used as reference to calculate the absolute reflectivity. 

The real part of the optical conductivities $\sigma_{1}(\omega)$ were obtained from the reflectivity data by means of Kramers-Kr\"{o}nig transformations. In order to perform this calculation, the reflectivity was simulated by a Lorentz oscillator centered around 24000 cm$^{-1}$ in the high-frequency range, and set at a constant value between 0 and 50 cm$^{-1}$. We verified that such an assumption produces no noticeable effect on the conductivity spectra presented below.

LDC were carried out by means of a shell model using the general utility lattice program (GULP)\cite{Gale97}.

\section{Results and discussion}

\subsection{Spectra of PrMnO$_3$ and CaMnO$_3$}

\begin{figure}
\includegraphics[scale=1]{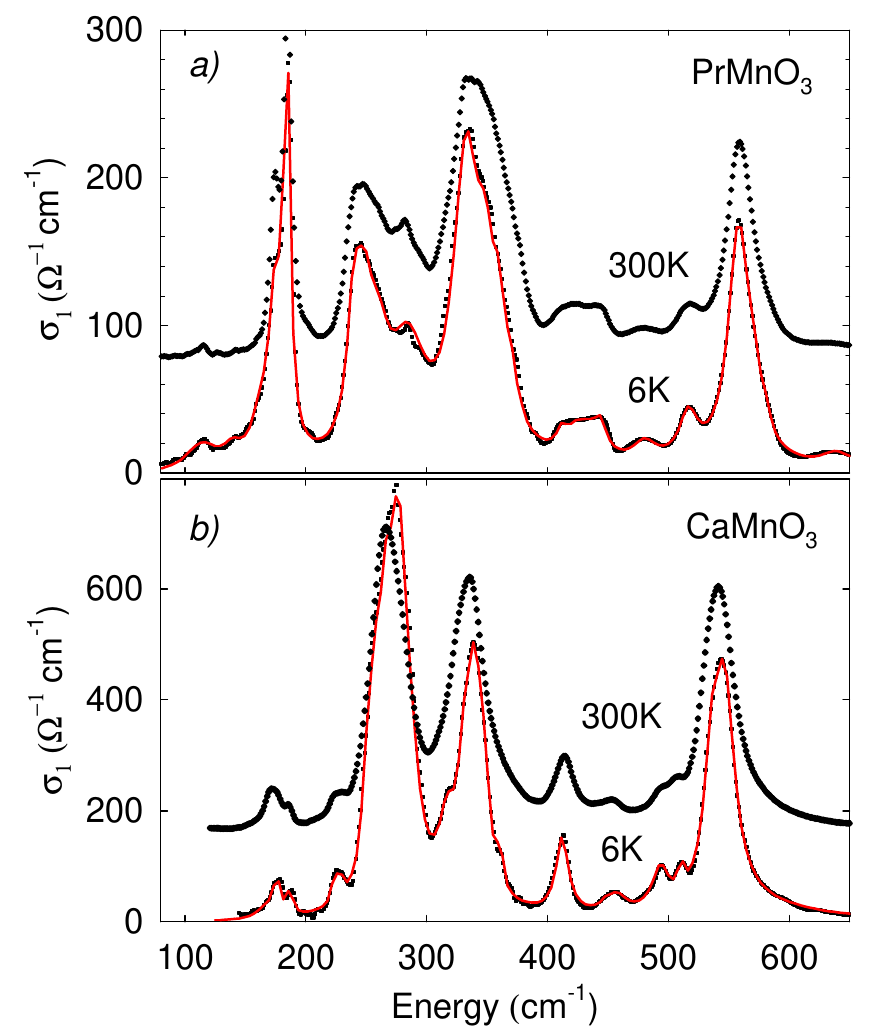}
\caption{(Color online) Infrared phonon modes in (a) PrMnO$_3$ and (b) CaMnO$_3$, at 6 K and 300 K. The red solid lines represent the best fit of Eq.\,\ref{equation1} to the low-temperature data. For better clarity, the conductivity values have been shifted by (a) 75 $\Omega^{-1}$cm$^{-1}$ and (b) by 150 $\Omega^{-1}$cm$^{-1}$, for the room temperature curves.}
\label{figure1}
\end{figure}

Figs.\,\ref{figure1}(a) and 1(b) respectively display the real part of the optical conductivity of PrMnO$_3$ and CaMnO$_3$ at $300$ and $6$ K in the far-IR frequency range which corresponds to the region of optical phonon vibrations. These spectra are consistent with the optical conductivity typical of an insulator that shows several peaks associated with the IR-active transverse-optical modes. As the temperature is lowered, no appreciable spectral change is observed, and the phonon features become narrower and sharper, as it is usually expected.

The spectra shown in Figs.\,\ref{figure1}(a) and 1(b) appear to be rather qualitatively similar in structure. They exhibit, around 160, 300, and 550 cm$^{-1}$, three main phonon groups showing a fine structure. In particular, it may be seen that the phonon group located between 250 cm$^{-1}$ and 375 cm$^{-1}$ --- known as the bending band --- is split for both PrMnO$_3$ and CaMnO$_3$, in agreement with what is expected in orthorhombically distorted perovskites with \textit{Pnma} structure \cite{Couzi72}. The fine structure in this band is expected to be very sensitive to basic distorsions $D_{\lbrack101\rbrack}$ and $D_{\lbrack010\rbrack}$ associated respectively to tilts of the MnO$_6$ octahedra around the $\lbrack101\rbrack$ and $\lbrack010\rbrack$ cubic axes. So, since $D_{\lbrack101\rbrack}$ ($D_{\lbrack010\rbrack}$) in PrMnO$_3$ is 1.26 (1.34) times larger than in CaMnO$_3$ \cite{Abrashev02, Iliev06a}, the fine structure of the bending band in CaMnO$_3$ should appear significantly reduced compared to that of PrMnO$_3$. Indeed, such a general trend is clearly seen in Figs.\,\ref{figure1}(a) and 1(b).
 
On the other hand, the high-energy modes (centered around $\sim 550$ cm$^{-1}$) are thought to involve mainly stretching vibrations of MnO$_6$ octahedra. The frequencies of these modes are therefore expected to be directely related to the force constants of Mn--O bonds, which are determined by the interatomic distances. Taking into account the much shorter Mn--O distances in CaMnO$_3$ compared to PrMnO$_3$ \cite{Jirak97a, Alonso00, Blasco00, Martin02}, one could expect the peak positions of the stretching modes to satisfy the inequality  $\omega_{CMO}>\omega_{PMO}$. As seen from Figs.\,\ref{figure1} and \ref{figure2}, this is obviously not the case. It is worthy of note that the high-energy IR modes of LaMnO$_3$ --- in which the Mn--O distances are also greater than those in CaMnO$_3$ --- were also observed at higher frequencies than those of CaMnO$_3$ in a previous study of the IR absorption spectra \cite{Fedorov99}. Such a behaviour can be explained by the fact that the charge replacement of Mn$^{3+}$ in PrMnO$_3$ (LaMnO$_3$) with Mn$^{4+}$ in CaMnO$_3$ is accountable for the decrease of the frequencies of these high-energy IR modes.
  
The low-energy phonon group --- corresponding to external modes --- mainly involves $R=$ Pr or Ca atoms. Therefore, the frequencies of these modes depend on the mass $M_{R}$ of the atom $R$, and are sensitive to force constants of $R$--O bonds, which are determined by the $R$--O distances $d_{R-O}$. In an harmonic oscillator approximation ($\omega=\sqrt{k/M_{R}}$), assuming a spring constant $k$ governed by $1/\langle d_{R-O} \rangle^{3}$ dependence \cite{Atanassova94}, and using the values of mean R--O distances available in Refs.\,\onlinecite{Alonso00} and \onlinecite{Martin02}, the external mode frequencies in CaMnO$_3$ and PrMnO$_3$ should scale as
$\omega_{CMO}/\omega_{PMO}\simeq1.9$. Figs.\,\ref{figure1} and \ref{figure2} obviously show that this is not in quantitative agreement with the experimental data. So, such a simple interpretation in terms of harmonic oscillator is inadequate to explain the frequency shift of these modes, and one has to take into account the more complex orthorhombic phonon vibrations. Finally, a comparison of Figs.\,\ref{figure1}(a) and 1(b) reveals very strong attenuation of the amplitude of some of these modes in CaMnO$_3$.
 
It is worthy of note that the present data which show a high number of observed phonons for CaMnO$_3$ and some similarity above $\sim 220$ cm$^{-1}$ between the spectra of PrMnO$_3$ and CaMnO$_3$, agree with the above mentioned study of the IR absorption spectra of LaMnO$_3$ and CaMnO$_3$ by Fedorov \textit{et al.} \cite{Fedorov99}. The existence of such a correlation between the spectra of these perovskites shows that the $D_{JT}$ basic lattice distorsion --- which originates from the JT effect in the case of PrMnO$_3$ and LaMnO$_3$ --- is not the driving force of the orthorhombic distorsion in both these materials since their respective $D_{JT}$ distorsion is almost 60 times larger than in CaMnO$_3$\cite{Abrashev02, Iliev06a}.    

\subsection{Oscillators model Fit}

To examine in some detail the phonon modes, we have fitted the spectra shown in Figs.\,\ref{figure1}(a) and 1(b) using the well known Lorentz model: 
\begin{eqnarray}	
\sigma_{1}(\omega)=\sum_{j}\frac{S_{j}\gamma_{j}\omega^{2}}{(\omega_{j} ^{2}-\omega^{2})^{2}+\gamma_{j}^{2}\omega^{2}}
\label{equation1}
\end{eqnarray}
where $S_{j}$, $\gamma_{j}$, and $\omega_{j}$ are respectively the amplitude, width, and frequency of the \textit{j}th oscillator. The red solid lines in Figs.\,\ref{figure1} display the best fit obtained using the minimum number of oscillators that allow to achieve a reliable fit to the spectra at low temperature. It is convenient to define the normalized amplitude $\tilde{S}_{j}=100\times S_{j}/\sum S_{j}$ that corresponds to the percentage of spectral weight associated with the \textit{j}th oscillator. We list in Table \ref{tab:table1} the values of the fitting parameters.

\begin{table*}
\caption{\label{tab:table1} Best-fit values of the normalized amplitudes ($\tilde{S}_{j}$), widths ($\gamma_{j}$), and  frequencies ($\omega_{j}$) of the low temperature spectra shown in Figs.\,\ref{figure1}(a) and 1(b). The frequency and the symetry of the different phonon modes obtained by means of the lattice dynamics calculations (LDC) are also reported. $\omega_{j}$ and $\gamma_{j}$ are in cm$^{-1}$, $\tilde{S}_{j}$ is in $\%$.} 
\begin{ruledtabular}
\begin{tabular}[b]{c c c c c c c c c c}
\multicolumn{5}{c}{PrMnO$_3$} & \multicolumn{5}{c}{CaMnO$_3$} \\
\multicolumn{3}{c}{Best fit} & \multicolumn{2}{c}{LDC} & \multicolumn{3}{c}{Best fit} & \multicolumn{2}{c}{LDC} \\
$\tilde{S}_{j}$ & $\gamma_{j}$ & $\omega_{j}$ & Symmetry & $\omega_{j}$ &  $\tilde{S}_{j}$ & $\gamma_{j}$ & $\omega_{j}$ & Symmetry & $\omega_{j}$\\
\hline
1.6 & 21 & 115 & $B_{3u}$ & 125  &  & &  & $B_{2u}$ & 133   \\
0.7 & 17 & 141 & $B_{1u}$ & 132 & 0.6 & 17 & 166 & $B_{3u}$ & 172 \\
2.2 & 22  & 162 & $B_{2u}$ & 148  & 1.1 & 8 &  176 & $B_{3u}$ & 181 \\
6.1 & 13 & 175 & $B_{1u}$ & 184  & 0.7 & 8 & 187 & $B_{1u}$ & 190 \\
6.9 & 6 & 184 & $B_{3u}$ & 197  & 0.8 & 8 & 227 & $B_{1u}$ & 224 \\
7.6 & 17 & 244 & $B_{2u}$ & 241 &  &  &  & $B_{2u}$ & 248.5 \\   
9.2 & 26 & 257 & $B_{1u}$ & 279 & 5.6 & 16 & 254 & $B_{3u}$ & 250 \\
7.6 & 28 & 283 & $B_{3u}$ & 298 & 20.0 & 18 & 276 & $B_{2u}$ & 275 \\ 
1.2 & 24 & 298 & $B_{3u}$ & 301 &  &  &  & $B_{3u}$ & 281 \\
20.9 & 24 & 333 & $B_{1u}$ & 348 & 16.5 & 20 & 266 & $B_{1u}$ & 284 \\
7.1 & 18 & 349 & $B_{2u}$ & 351.5  & 2.2 & 12 & 286 & $B_{1u}$ & 298 \\
0.9 & 16 & 369 & $B_{1u}$ & 370 & 3.9 & 16 & 317 & $B_{2u}$ & 324 \\
3.8 & 16 & 360 & $B_{2u}$ & 379 & 8.7 & 13 & 340 & $B_{3u}$ & 345 \\
 &  &  & $B_{3u}$ & 387 & 6.4 & 14 & 332 & $B_{1u}$ & 346.5 \\
0.6 & 10 & 412 & $B_{3u}$ & 396.5 & 1.1 & 13 & 361 & $B_{3u}$ & 364 \\
1.1 & 17 & 423 & $B_{1u}$ & 403 & 2.2 & 12 & 348 & $B_{2u}$ & 368 \\
1.2 & 18 & 433.5 & $B_{2u}$ & 421 & & & & $B_{1u}$ & 379.5 \\
1.1 & 12 & 444 & $B_{1u}$ & 426 & 3.3 & 12 & 412 & $B_{1u}$ & 414 \\
1.3 & 22 & 480 & $B_{3u}$ & 486 & 0.2 & 14 & 449 & $B_{2u}$ & 450 \\
1.8 & 14 & 515 & $B_{1u}$ & 517.5 & 1.2 & 17 & 457.5 & $B_{3u}$ & 460 \\
0.9 & 14 & 524 & $B_{2u}$ & 525 & 2.2 & 14 & 494 & $B_{1u}$ & 496 \\
2.8 & 17 & 557 & $B_{3u}$ & 555 & 2.1 & 15 & 510 & $B_{2u}$ & 504 \\
10.5 & 18 & 558 & $B_{2u}$ & 559 & 9.2 & 15 & 539 & $B_{3u}$ & 534 \\
2.4 & 16 & 571 & $B_{1u}$ & 562 & 10.5 & 16 & 548 & $B_{1u}$ & 544 \\
0.5& 10 & 580 & $B_{3u}$ & 584 & 1.5 & 36 & 584 & $B_{3u}$ & 613 \\ 
\end{tabular}
\end{ruledtabular}
\end{table*}
 
\subsection{Lattice dynamical calculations}

In order to identify the observed phonon peaks, LDC for both PrMnO$_3$ and CaMnO$_3$ have been performed by means of a shell model using GULP\cite{Gale97}. In this model, each ion is represented by a massless shell of charge $Y$ and a core of charge $X$ which are coupled by an harmonic spring constant $K$. The interionic interactions are described by long range Coulomb potentials and short range potentials of the Born-Mayer-Buckingham form: $$V(r)=A\:exp(-r/\rho)-C/r^{6}$$ where $r$ is the interatomic distance. These two-body radial potentials are of course a first approximation in which the effect of non-spherical potential terms of the interactions and other corrections have been neglected. However, it was shown that LDC using a shell model with this type of interionic interaction potential yield valuable information about the phonon eigenmodes of various oxydes and can be used to perform the assignment of their experimental phonon spectra \cite{Popov95}.

Starting from experimental crystallographic data, the lattice energy is minimized with respect to unit cell variables --- namely, the lattice parameters and atomic coordinates --- while large potential parameter space zones are explored. We determine the potential parameters such that the lattice energy is minimum, and they reproduce the phonon frequencies close to those determined by Raman and IR spectroscopy experiments. The values of parameters $A$ and $\rho$ for the O--O short range interaction are taken from Refs.\,\onlinecite{Lewis85} and \onlinecite{Cherry95}, which have been successfully used for the modeling of oxides \cite{Popov95, Gale96}. For PrMnO$_3$ and CaMnO$_3$, we obtain the best fit with the parameters listed in Table \ref{tab:table2}. It is worthy of note that the values of the potential parameters for the Mn$^{3+}$ ions derived in the present work are very close to those obtained by Islam \textit{et al.} \cite{Islam96}.

\begin{table*}
\caption{\label{tab:table2} Sell model parameters. Cut-off radii of 10 \AA \ and 12 \AA \ were respectively used for the cation-oxygen and oxygen-oxygen short range potentials.}
\begin{ruledtabular}
\begin{tabular}{l c c c c c c c}
Ion & $X (\vert e\vert)$ & $Y (\vert e\vert)$ & $K$ (eV \AA$^{-2}$) & Ionic pair & $A$ (eV) & $\rho (\AA)$ & $C$ (eV \AA$^{-6}$) \\
\hline
Pr & 1.5 & 1.5 & 30 & Pr--O & 45461 & 0.2272 & 0 \\
Ca & $-$0.6 & 2.6 & 80 & Ca--O & 10040 & 0.2401 & 0 \\
Mn$^{3+}$ & 0.5 & 2.5 & 95 & Mn$^{3+}$--O & 1337 & 0.3260 & 0 \\
Mn$^{4+}$ & 1.9 & 2.1 & 40 & Mn$^{4+}$--O & 4980 & 0.2650 & 0 \\
O & 0.24 & $-$2.24 & 42 & O--O & 22764 & 0.1490 & 27.88 \\
\end{tabular}
\end{ruledtabular}
\end{table*}

To check the validity of our LDC, we show in Table \ref{tab:table3} a comparison between the calculated structural parameters and those observed experimentally \cite{Cherepanov, Jirak97a, Alonso00, Zhou06, Blasco00, Jorge}. It can be inferred that the LDC performed here produce a reasonable agreement with the experimental crystallographic data. In particular, the calculated lattice constants are very close to the observed values with a discrepancy of less than 2\%. On the other hand, in accordance with the results of the group theory for the $\Gamma$-point of the orthorhombic $Pnma$ structure of $R$MnO$_3$ \cite{Couzi72, Iliev98, Smirnova99}, we obtain 60 vibration modes of which 24 ($7\:A_{g}+5\:B_{1g}+7\:B_{2g}+5\:B_{3g}$) are Raman active, 25 ($9\:B_{1u}+7\:B_{2u}+9\:B_{3u}$) are IR-active, 8 ($8\:A_{u}$) are silent, and 3 ($B_{1u}+B_{2u}+B_{3u}$) are acoustic. For both PrMnO$_3$ and CaMnO$_3$, we find good agreement between the calculated Raman-active phonon modes and the available experimental Raman spectra \cite{Abrashev02, Iliev06a, Laverdiere06a, Laverdiere06b, Martin03, Martin02, Granado01, Liarokapis99, Iliev03}. In particular, the average deviation between the calculated and observed Raman frequencies is less than 10\%. Indeed, the two dominant peaks observed by Raman spectroscopy in PrMnO$_3$ around 605 and 485 cm$^{-1}$ and assigned to the symmetric [$B_{2g}(1)$] and asymmetric  [$A_{g}(1)$] stretching of O2 ions in the $xz$ planes \cite{Iliev06a, Laverdiere06a, Laverdiere06b, Martin03, Martin02} are respectively predicted by the present calculation at 610 and 466 cm$^{-1}$ (the numbering of the modes follows Refs.\,\onlinecite{Iliev06a, Laverdiere06a, Laverdiere06b, Abrashev02}, and \onlinecite{Iliev98}).
Moreover, the other main Raman peaks observed in PrMnO$_3$ around 465, 450 and 325 cm$^{-1}$ which have been respectively assigned \cite{Iliev06a, Laverdiere06a, Laverdiere06b, Martin03, Martin02} to the MnO$_6$ bending [$A_{g}(3)+B_{2g}(3)$] and tilt [$A_{g}(4)$] modes, are reproduced by our calculation at frequencies 510, 508 and 357 cm$^{-1}$, respectively. In the CaMnO$_3$ case, it is admitted \cite{Abrashev02, Martin02} that the two pairs of Raman peaks observed around 487 and 466 cm$^{-1}$ and around 258 and 245 cm$^{-1}$ cannot be related to $A_{g}(1)+B_{2g}(1)$ (stretching) modes --- the dominant ones in PrMnO$_3$. The Raman frequencies of the doublet around 487 and 466 cm$^{-1}$ which has been assigned \cite{Abrashev02, Martin02} to $A_{g}(3)+B_{2g}(3)$ (bending) modes, were obtained from a previous shell model computation by Abrashev \textit{et al.} \cite{Abrashev02} at 467 and 453 cm$^{-1}$, whereas the present calculation gives these frequencies at 510 and 503 cm$^{-1}$. Finally, it should be emphasized that the assignment of the pair around 258 and 245 cm$^{-1}$ still remains controversial. By comparison between observed and calculated Raman frequencies \cite{Abrashev02}, this doublet was related to $A_{g}(7)+B_{2g}(7)$ modes involving the displacements of the apical oxygen (O1) and Ca atoms, while --- based on the correlation between frequencies and relevant structural parameters of La$_{1-x}$Ca$_x$MnO$_3$ series --- Mart\'{i}n-Carr\'{o}n \textit{et al.} \cite{Martin02} assigned it  to $A_{g}(4)+B_{2g}(4)$ (tilt) modes. 

\begin{table*}
\caption{\label{tab:table3} Comparison of the lattice parameters of PrMnO$_3$ and CaMnO$_3$ observed experimentally at room temperature ($Pnma$ space group) with those obtained by means of the lattice dynamics calculations (LDC). The last seven lines represent the fractional atomic coordinates; $R$ (Pr$/$Ca), Mn, apical (O1) and basal (O2) oxygen atoms are respectively located at Wyckoff positions 4$c$ ($x,1/4,z$), 4$b$ ($0,0,1/2$), 4$c$ ($x,1/4,z$) and 8$d$ ($x,y,z$), and at their symmetry equivalent positions. }
\begin{ruledtabular}
\begin{tabular}{r c c c c c c c c c}
 & \multicolumn{5}{c}{PrMnO$_3$} & \multicolumn{4}{c}{CaMnO$_3$} \\
 & Observed\footnotemark[1] & Observed\footnotemark[1] & Observed\footnotemark[1] & Observed\footnotemark[2] & LDC & Observed\footnotemark[2] & Observed\footnotemark[2] & Observed\footnotemark[2] & LDC \\
& Ref. \onlinecite{Cherepanov}& Ref.\,\onlinecite{Jirak97a} & Ref.\,\onlinecite{Alonso00} & This work &  & Ref.\,\onlinecite{Zhou06} & Ref.\,\onlinecite{Blasco00} & Ref.\,\onlinecite{Jorge} &  \\
\hline
$a$ (\AA)& 5.5914 & 5.786 & 5.8129 & 5.7852 & 5.6658 & 5.28287 & 5.2813 & 5.279 & 5.3110 \\
$b$ (\AA)& 7.672 & 7.589 & 7.5856 & 7.5939 & 7.7042 & 7.45790 & 7.4582 & 7.460 & 7.4454 \\
$c$ (\AA)& 5.4562 & 5.450 & 5.4491 & 5.4490 & 5.4739 & 5.26746 & 5.2687 & 5.273 & 5.2446 \\
$V$ (\AA$^{3}$)& 234.05 & 239.31 & 240.278 & 239.386 & 238.944 & 207.5332 & 207.528 & 207.657 & 207.388 \\
R (4c) $x$ & 0.0451 & 0.066 & 0.0639 & 0.062 & 0.0517 & 0.0330 & 0.0313 & 0.030 & 0.0320 \\
$z$ & 0.9931 & 0.983 & 0.9911 & 0.991 & 0.9913 & 0.9944 & 0.9929 & 0.995 & 0.9924  \\
O1 (4c) $x$ & 0.4825 & 0.484 & 0.4819 & 0.482 & 0.4708 & 0.4990 & 0.4869 & 0.491 & 0.4701\\
$z$ & 0.0770 & 0.084 & 0.0834 & 0.105 & 0.0757 & 0.0657 & 0.0624 & 0.066 & 0.0782\\
O2 (8d) x & 0.2978 & 0.313  & 0.3174 & 0.311 & 0.2920 & 0.2870 & 0.2882 & 0.288 & 0.2876\\
$y$ & 0.0404 & 0.042 & 0.0430 & 0.042 & 0.0404 & 0.0332 & 0.0342 & 0.033 & 0.0417\\
$z$ & 0.7156 & 0.716 & 0.7151 & 0.730 & 0.7058 & 0.7116 & 0.7118 & 0.709 & 0.7112 \\
\end{tabular}
\end{ruledtabular}
\footnotetext[1]{by neutron diffraction}
\footnotetext[2]{by X-ray diffraction}
\end{table*}    

\subsection{Assignment of infrared-active phonon modes}

\begin{figure}
\includegraphics[scale=1]{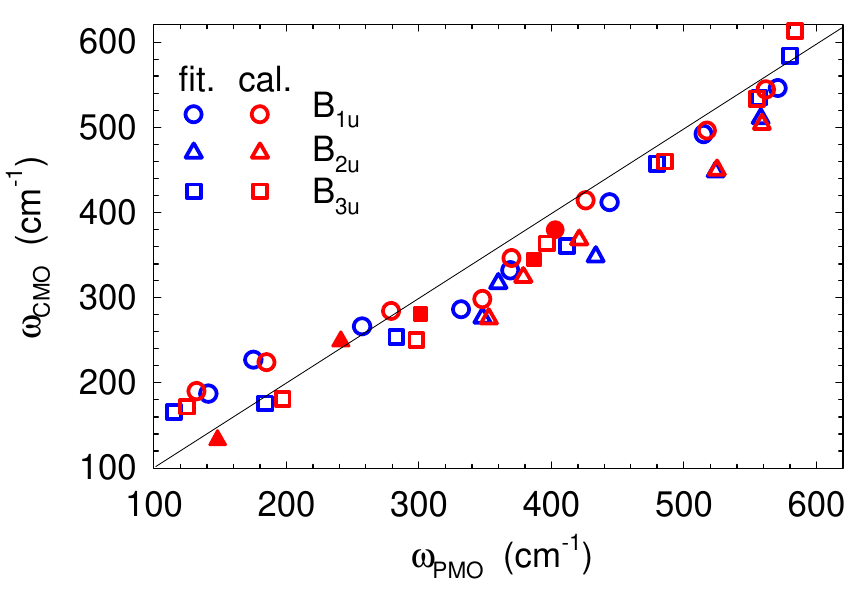}
\caption{(Color online) Correspondence between the fitted (calculated) infrared active-phonon frequencies of PrMnO$_3$ [abscissae axis] and those of CaMnO$_3$ [ordinates axis]. The blue (red) symbols denote the fitted (calculated) frequencies; the meaning of filled symbols is given in the text, and the straight line is a guide for the eyes. Since the infrared spectra are obtained from polycrystalline samples, the assignment of experimental peaks to $B_{1u}$, $B_{2u}$ and $B_{3u}$ modes is made by comparing the calculated frequencies with those determined from the best fit of the Eq.\,\ref{equation1} to the data.}
\label{figure2}
\end{figure}

As our measurements were done on polycrystalline samples, the IR-active phonon peaks corresponding to modes of $B_{1u}$, $B_{2u}$ and $B_{3u}$ symmetry are undistinguishable. For this reason, we make the assignment in order that the average deviation between the calculated and fitted phonon frequencies is minimum. The results of this comparison are shown in Table \ref{tab:table1}. It can be noted that very few ambiguities about the assignment of phonons remain, such as, for instance, the two peaks in the spectrum of PrMnO$_3$ at 360 and 369 cm$^{-1}$ which could be equally related to the $B_{1u}$ and $B_{2u}$ modes at 370 and  379 cm$^{-1}$, respectively. In such cases, the assignment of the peaks is obtained by comparing the experimental amplitudes to those estimated qualitatively within the framework of a rigid-ion approach, as described below. Taking into account the assignment of the modes and representing the fitted (calculated) values of $\omega_j$ by blue (red) symbols, we show in Fig.\,\ref{figure2} the relationship between the IR phonon frequencies of PrMnO$_3$ and CaMnO$_3$. Note that the experimental frequencies at 162, 244, 283, and 423 cm$^{-1}$ for PrMnO$_3$ do not appear since they have no counterpart in the fitted spectrum of CaMnO$_3$, and vice versa for the modes at 340 cm$^{-1}$ for CaMnO$_3$ (in all these cases, the calculated frequencies are represented by filled symbols in the figure). It is clear from Fig.\,\ref{figure2} that the IR phonon frequencies observed in PrMnO$_3$ and CaMnO$_3$ are closely correlated, and that this result is quantitatively in good agreement with the present LDC. On the other hand, we also made a comparison between the fitted values of $S_j$ for each IR phonon mode of PrMnO$_3$ and CaMnO$_3$, and no correlation between them has been found. This suggests that the replacement of the Pr$^{3+}$ ions with Ca$^{2+}$ drastically changes the vibrational patterns of IR phonons.

\subsection{ Analysis of the infrared active modes}

\begin{figure}
\includegraphics[scale=1]{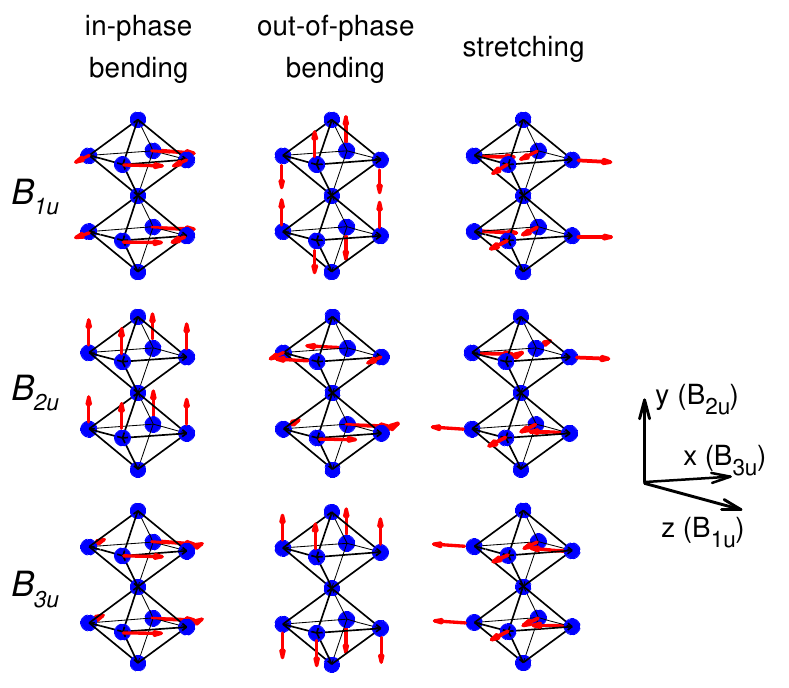}
\caption{(Color online) Classification of vibrations of basal oxygen (O2) atoms for each mode of symmetry  $B_{1u}$, $B_{2u}$, and $B_{3u}$. Because the unit cell of $R$MnO$_{3}$ compounds with the $Pnma$ structure contains two MnO$_{6}$ octahedra along the $y$ direction, the O2 atoms in these neighboring octahedra may vibrate in-phase or out-of-phase.}
\label{figure3}
\end{figure}

Let us briefly remember some general considerations about the expected forms of the IR-active modes \cite{Kroumova} for discussing --- on the basis of the present assignment --- the atomic vibrations related to the phonon peaks observed in PrMnO$_3$ and CaMnO$_3$. Since the Pr/Ca and O1 atoms in the $Pnma$ space group occupy Wyckoff position 4$c$, their motions are restricted by the site symmetry ($C_{s}^{xz}$) within the $xz$ planes for the $B_{1u}$ and $B_{3u}$ modes, and along the $y$ direction for the $B_{2u}$ mode. As for the Mn and O2 atoms --- which are respectively localized at Wyckoff positions 4$b$ (or 4$a$) and 8$d$ --- there are no symmetry restrictions on the direction of their vibrations for all IR-active modes. Nevertheless, we can show using the results of the group theory that the general movement of O2 atoms can be reduced for each mode of symmetry $B_{1u}$, $B_{2u}$, and $B_{3u}$ to a combination of three independent displacements involving in-phase or out-of-phase vibrations along (stretching) or perpendicular (bending) to the Mn-O2 bonds. The shape of these O2 vibrations is shown in Fig.\,\ref{figure3}. We see that the $B_{1u}$, $B_{2u}$ and $B_{3u}$ bending vibrations of O2 atoms can be either in-phase or out-of-phase, while the $B_{1u}$ and $B_{3u}$ stretching vibrations are in-phase and those with symmetry $B_{2u}$ are out-of-phase. Moreover, it can be noted that the $B_{1u}$ and $B_{3u}$ in-phase bending vibrations of O2 atoms are related to one another by a rotation of $90^{o}$ of MnO$_{6}$ octahedra around the $y$ axis. The same occurs with the $B_{1u}$ and $B_{3u}$ stretching vibrations of O2 atoms.

\begin{figure*}
\includegraphics[scale=1]{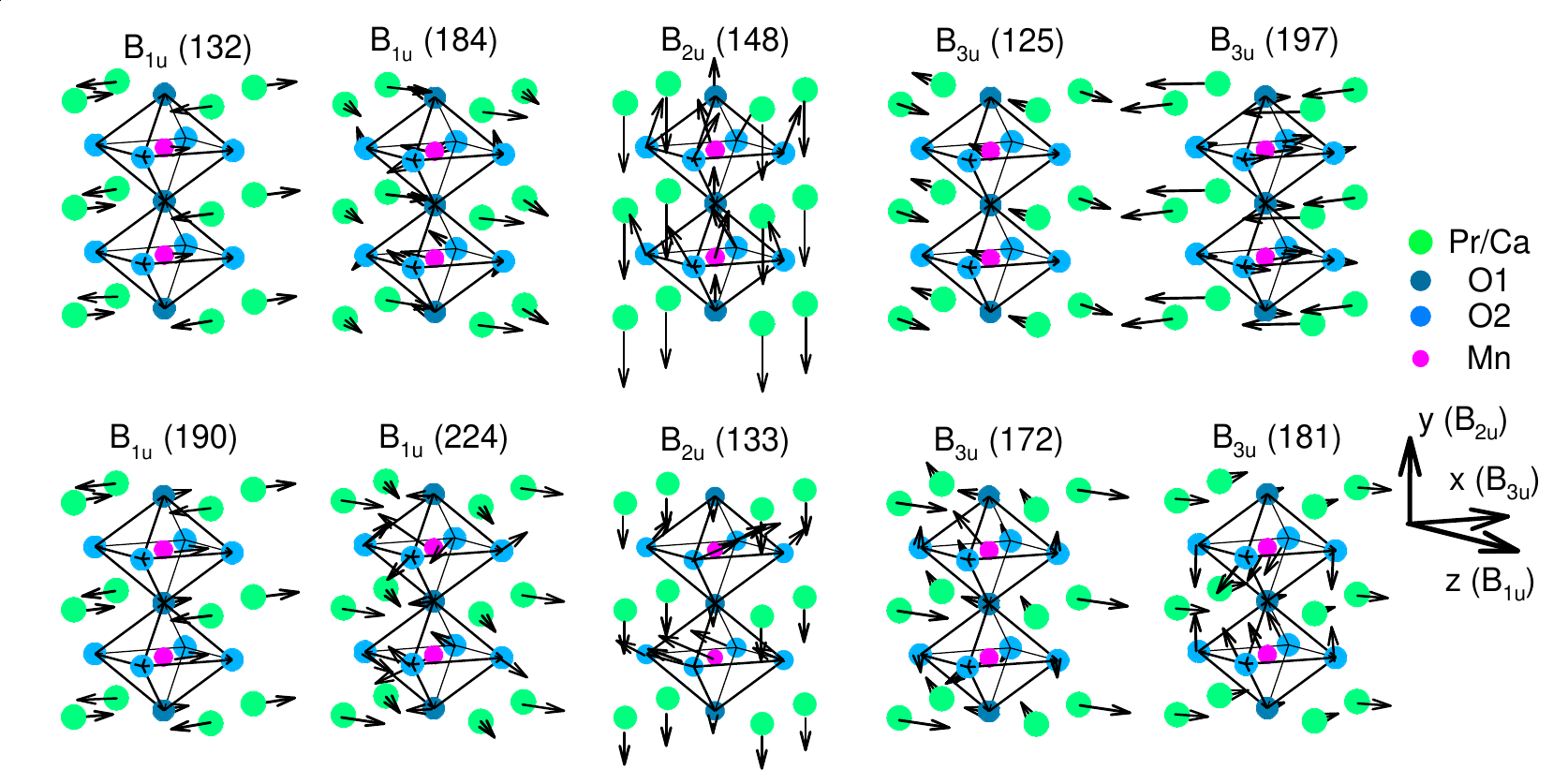}
\caption{(Color online) Atomic displacements of low-energy modes of $B_{1u}$, $B_{2u}$, and $B_{3u}$ symmetry of PrMnO$_3$ (top) and CaMnO$_3$ (bottom) with $Pnma$ structure. The numbers in brackets correspond to the calculated frequencies in cm$^{-1}$.}
\label{figure4}
\end{figure*}

Taking into account the effects of the atomic polarization, an expression for $S_j$ is given in Refs.\,\onlinecite{Chen92} and \onlinecite{Koval95}. In particular, for every orthorhombic crystal one finds that $S_j=\frac{1}{\Omega}\:\left|p_\alpha(j)\right|^2$ where $\Omega$ is the volume of the unit cell, and $p_\alpha(j)$ is the nonzero component of the electric dipole moment vector of the $j$th IR-active mode, namely $p_z$ for $B_{1u}$, $p_y$ for $B_{2u}$, and $p_x$ for $B_{3u}$ in the $Pnma$ coordinate axes orientation. If the shell vibrational contribution is neglected, then it can be shown \cite{Chen92, Koval95, Born} that $p_\alpha(j)=\sum_{\kappa} eZ_\kappa \: u_\kappa^\alpha(j)$ where the subscript $\kappa$ labels the atoms in the unit cell, $e Z_k$ is the effective charge of the $\kappa$th atom, and $u_{\kappa}^{\alpha}(j)$ denotes its active displacement component in the $j$th mode, so that $S_j= \frac{e^2}{\Omega}\lbrack \sum_{\kappa} Z_\kappa \; u_{\kappa}^{\alpha}(j)\rbrack ^2$. For instance, as can be seen in Fig.\,\ref{figure3}, the out-of-phase bending vibrations of O2 atoms (for all symmetries) and the stretching vibrations of O2 atoms with $B_{2u}$ symmetry produce a zero dipole moment. Of course, such a formula for $S_j$ derived from the rigid-ion model is very rough. However, for our purposes an important feature of this result is to highlight the  possibility that some IR-active modes characterized by relatively important atomic vibrations have a small resultant dipole moment, and hence a small oscillator strength $S_j$. As seen in Fig.\,\ref{figure4}, the low-energy modes of CaMnO$_3$ are the examples of such modes. Indeed, each of these modes does not produce a large dipole moment. For instance, the $z$ component of atomic displacements giving the dipole moment in the B$_{1u}$ modes is small for the one at 187 (190) cm$^{-1}$ (in the following, the numbers in brackets indicate the calculated frequencies) and is not represented in Fig.\,\ref{figure4}. The same occurs with the $x$ component for the $B_{3u}$ modes at 166 (172) and 176 (181) cm$^{-1}$. The B$_{1u}$ and B$_{2u}$ modes at 227 (224) and (133) cm$^{-1}$ provide another example of modes carrying a small dipole moment. For these modes, the active components of the Ca and Mn vibrations are relatively large but in the opposite direction while those of the O1 and O2 atoms are small, which leads to a small resultant dipole moment. In PrMnO$_{3}$, the situation is clearly different with the low-energy IR-active modes. At first glance, the B$_{1u}$ and B$_{2u}$ modes at 175 (184) and 162 (148) cm$^{-1}$ are respectively similar to those obtained at 227 (224) and (133) cm$^{-1}$ in CaMnO$_{3}$ but the active components of the O1 and O2 vibrations are large and in the same direction, and hence, the dipole moment of both these modes is not small. Furthermore, we also find that the dipole moment of the B$_{3u}$ mode at 184 (197) cm$^{-1}$ is not small due to a large $x$ component of atomic displacements. So, our calculations predict that unlike PrMnO$_{3}$, none of the low-energy IR-active modes of CaMnO$_{3}$ has a large oscillator strength. It can be clearly seen from Fig.\,\ref{figure1} that this prediction is in reasonably good agreement with the experimental data.

\begin{figure}
\includegraphics[scale=1]{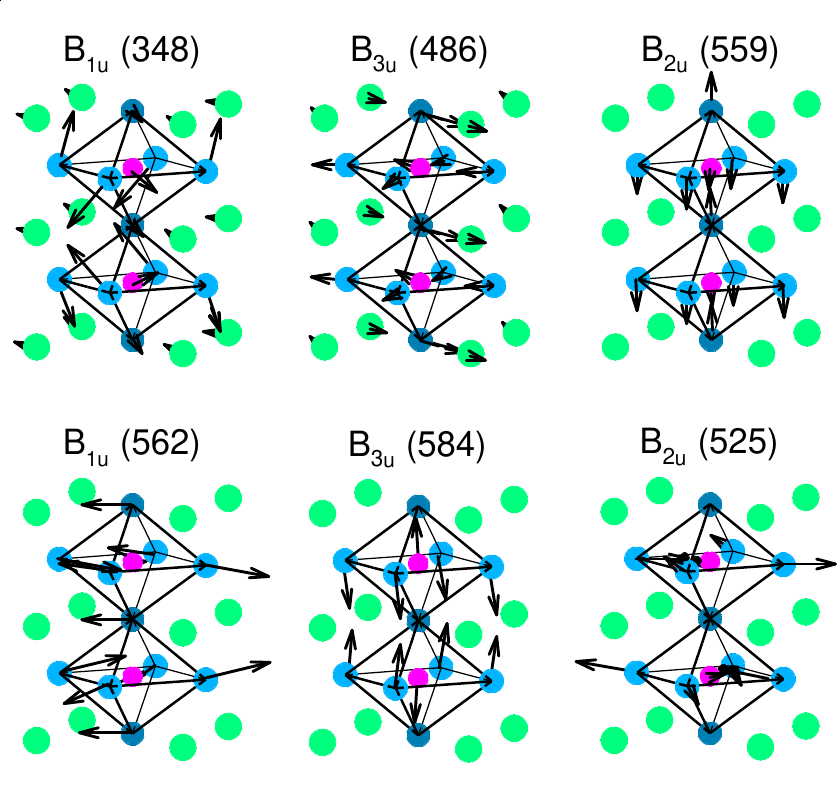}
\caption{(Color online) Movements of atoms for a selection of $B_{1u}$, $B_{2u}$, and $B_{3u}$ modes of PrMnO$_3$ with $Pnma$ structure. The symbols and the direction of the axes are defined in Fig.\,\ref{figure4}.}
\label{figure5}
\end{figure}

Fig.\,\ref{figure5} shows a representative selection of phonon modes of PrMnO$_3$ involving almost no movement of Pr atoms, as expected for internal modes. As can be seen, the modes at 333 (348), 480 (486), and 558 (559) cm$^{-1}$ --- as well as those at 283 (298), (387), 412 (396.5), 423 (403), and 515 (517.5) cm$^{-1}$ --- correspond to complex motions of MnO$_{6}$ octahedra for which the amplitudes of the vibrations of Mn atoms are comparable with those of oxygen atoms. For the mode at 580 (584) cm$^{-1}$ (see Fig.\,\ref{figure5}) --- as well as those at 360 (379), 433.5 (421), and 444 (426) cm$^{-1}$ --- only the O2 and Mn atoms play a role, so that the O1 atoms can be assumed to be at rest. On the other hand, there are also modes where only the oxygen atoms vibrate. In the mode at 571 (562) cm$^{-1}$ displayed in Fig.\,\ref{figure5} --- as well as the 557 (555) cm$^{-1}$ mode --- the O1 and O2 atoms vibrate with comparable amplitudes, whereas for the mode at 524 (525) cm$^{-1}$ (see  Fig.\,\ref{figure5}) the vibrations of O2 atoms dominate. As was stated above and shown in Fig.\,\ref{figure3}, the general movement of O2 atoms is a combination of bending and stretching vibrations. The $B_{1u}$ mode at 571 (562) cm$^{-1}$ provides an example of such a mixture consisting of $\sim$80\% of stretching vibrations, $\sim$17\% of in-phase bending vibrations, and $\sim$3\% of out-of-phase bending vibrations. Similarly, the $B_{2u}$ mode at 524 (525) cm$^{-1}$ consists of $\sim$88\% of stretching vibrations, $\sim$11\% of out-of-phase bending vibrations, and less than 1\% of in-phase bending vibrations. As precisely this latter type of vibration produces the dipole moment, its small magnitude leads to a very weak intensity. It should be emphasized that this mode is also found by Smirnova \cite{Smirnova99} in its LDC of LaMnO$_3$ ($B_{2u}$ mode at the frequency of 562 cm$^{-1}$). On the other hand, there are also modes for which the movement of O2 atoms correspond to either quasi-pure bending vibrations or quasi-pure stretching vibrations. So, the modes at 480 (486), 558 (559), and 580 (584) cm$^{-1}$ can be classified as pure stretching, as pure in-phase bending, and as pure out-of-phase bending, respectively. It is interesting to remark that the in-phase bending $B_{2u}$ mode at 558 (559) cm$^{-1}$ --- which is characterized by a large dipole moment resulting from displacements of oxygen and Mn atoms along the $y$ axis --- is also obtained in the LDC of LaMnO$_3$ reported by Smirnova \cite{Smirnova99} ($B_{2u}$ mode at the frequency of 516 cm$^{-1}$). 

Another significant point is that the atomic displacement patterns of the $B_{1u}$ mode at 571 (562) cm$^{-1}$ (see Fig.\,\ref{figure5}) is linked to that of the $B_{3u}$ modes at 557 (555) cm$^{-1}$ by a rotation of $90^{o}$ around the $y$ axis. The same occurs for the $B_{1u}$ and $B_{3u}$ modes at 423 (403) and 412 (396.5) cm$^{-1}$. For both of these pairs, the frequency difference between the corresponding $B_{1u}$ and $B_{3u}$ modes is expected to be small, and very sensitive to orthorhombic distortions. The experimental values $\Delta f\approx11 - 14$ cm$^{-1}$ turn out to be large compared with the calculated values of $\sim 7$ cm$^{-1}$. This suggests that the present LDC does not fully account  for the strutural distortion effects in PrMnO$_3$.

\begin{figure}
\includegraphics[scale=1]{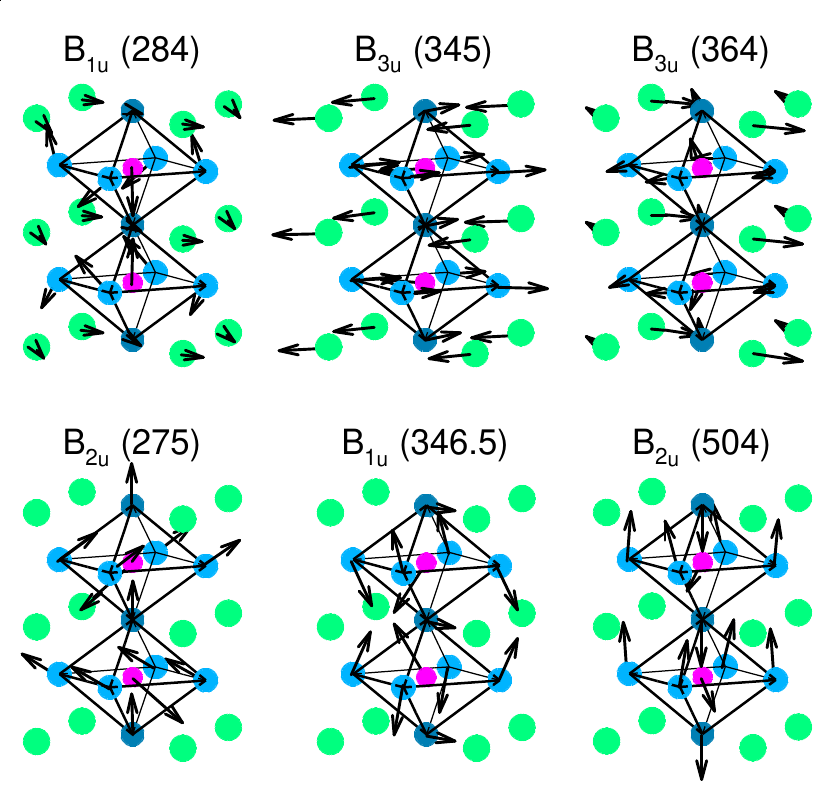}
\caption{(Color online) Movements of atoms for a selection of $B_{1u}$, $B_{2u}$, and $B_{3u}$ modes of CaMnO$_3$ with $Pnma$ structure. The symbols and the direction of the axes are defined in Fig.\,\ref{figure4}.}
\label{figure6}
\end{figure}

Fig.\,\ref{figure6} shows a representative selection of the phonon modes of CaMnO$_3$. As can be seen, the modes at 266 (284), 340 (345), and 361 (364) cm$^{-1}$, as well as those at (248.5), 254 (250), (281), 286 (298), 317 (324), 348 (368), (379.5), 412 (414), 449 (450), 457.5 (460), 539 (534), and 548 (544) cm$^{-1}$, involve both the movements of the Ca atoms and the MnO$_6$ octahedra. Therefore, the distinction between external and internal modes appears to be less clear in CaMnO$_3$ than in PrMnO$_3$. It follows that the Ca vibrations may participate significantly in the dipole moment of some of these modes. For instance, the dipole moment of the B$_{1u}$ mode at 266 (284) cm$^{-1}$ (see  Fig.\,\ref{figure6}) is assumed to be mainly caused by the $z$ component of the Ca vibrations and by the active O2 vibrations, namely the stretching and in-phase bending modes. Similarly, the B$_{3u}$ mode at 340 (345) cm$^{-1}$ (see  Fig.\,\ref{figure6}) --- in which all the atoms oscillate along the $x$ direction with comparable amplitudes --- is expected to produce a rather significant dipole moment, which should be enhanced by the Ca and Mn displacements in the opposite direction from that of the oxygen atoms. In contrast, the Ca motions do not play any essential role in the dipole moment induced by the other modes at (281), 286 (298), 361 (364), 539 (534), and 548 (544) cm$^{-1}$. In particular, in the B$_{3u}$ mode at 361 (364) cm$^{-1}$ (see Fig.\,\ref{figure6}) the modulus of calcium displacement vectors happens to be relatively large. However, their $x$ component responsible for the dipole moment is small. Finally, as shown in Fig.\,\ref{figure6} the modes at 276 (275), 332 (346.5), and 510 (504) cm$^{-1}$ --- as well as those at 494 (496), and 584 (613) cm$^{-1}$ --- mainly correspond to vibrations of the MnO$_6$ octahedra. So, these modes in which the Ca atoms can be considered to be at rest, can be classified as purely internal. They consist of complex motions of the Mn and O atoms, except for the two modes at 494 (496), and 584 (613) cm$^{-1}$ where the displacements of the O2 atoms correspond to nearly pure stretching vibrations ($\sim$85\%), and pure out-of-phase bending vibrations, respectively. Nevertheless, although the amplitudes of the displacements of the Mn and O atoms are comparable, the dipole moment of these modes should be mainly due to vibrations of the O atoms, apart for the two at 276 (275), and 332 (346.5) cm$^{-1}$ (see Fig.\,\ref{figure6}) where the Mn and O1 motions play similar roles. More generally, the dipole moments induced by the modes of frequency greater than 400 cm$^{-1}$ are mainly due to movements of the O atoms. For the mode at 412 (414) cm$^{-1}$ the O2 vibrations dominate. For those at 457.5 (460), 494 (496), 539 (534), and 548 (544) cm$^{-1}$ the motions of the O1 atoms play the main role. Only for the mode at 510 (504) cm$^{-1}$ (see Fig.\,\ref{figure6}) the active components of the displacements of the O1 and O2 atoms are large and comparable, but in the opposite direction.

It follows that the vibrational patterns of IR phonon modes in PrMnO$_3$ and CaMnO$_3$ should be quite different between them, except for the modes at high frequencies. In fact, among the latter, only the $B_{3u}$ modes at 580 (584) cm$^{-1}$ in PrMnO$_3$ (see Fig.\,\ref{figure5}) and 584 (613) cm$^{-1}$ in CaMnO$_3$ correspond to the same atomic displacements. Note that this phonon mode --- involving mainly mixed vibrations of Mn and O2 atoms along the $y$ axis (see Fig.\,\ref{figure5}) --- is also obtained in LaMnO$_3$ (at 506 cm$^{-1}$) by means of the LDC reported by Smirnova \cite{Smirnova99}. Therefore, it is plausible to expect that this $B_{3u}$ mode could be common to the $R$MnO$_3$ compounds with $Pnma$ structure. Note also that the expected small value for the oscillator strenght of this phonon mode is consistent with the experimental data. On the other hand, like the $B_{1u}$ and $B_{3u}$ modes at 571 (562) and 557 (555) cm$^{-1}$ in PrMnO$_3$, the vibrational patterns of the $B_{1u}$ and $B_{3u}$ modes at 548 (544) and 539 (534) cm$^{-1}$ in CaMnO$_3$ are also related to one another by a rotation of $90^{o}$ around the $y$ axis. As mentioned above, the frequency difference $\Delta f $ between the corresponding $B_{1u}$ and $B_{3u}$ modes can be seen as a measure of the deviation of the orthorhombic structure from the ideal (cubic) perovskite structure. Therefore, since the structural distorsion effects in PrMnO$_3$ are stronger than those in CaMnO$_3$, we expect to observe $\Delta f_{PMO}>\Delta f_{CMO}$, which is in good agreement with the experimental data. Finally, it should be noticed that the vibrational patterns of the $B_{1u}$ and $B_{3u}$ modes at 548 (544) and 539 (534) cm$^{-1}$ in CaMnO$_3$ involve small amplitudes for both of these modes, which is in poor agreement with the experimental data (see Table \ref{tab:table1}). However, taking into account the complexity of structural properties of these materials, we believe that the overall agreement between experiment and theory should be considered as reasonable.

\section{Conclusion}

We have studied the IR reflectivity spectra of polycrystalline orthorhombic PrMnO$_3$ and CaMnO$_3$, the parent compounds of charge ordering manganites. LDC has been performed using a shell model for the two materials. The calculations have been carried out while optimizing the model parameters in order to match the calculated structural parameters, Raman, and IR phonon frequencies with their experimental values. By making a comparative study of the experimental and calculated frequencies we have been able to find a reasonable assignment of the IR modes observed in PrMnO$_3$ and CaMnO$_3$. In agreement with the calculations, we find a close relationship between the IR phonon frequencies of the two materials. This result shows that, contrary to what might be expected in an harmonic oscillator approximation, the frequencies of medium- and high-energy phonons in PrMnO$_3$ are higher than in CaMnO$_3$. Nevertheless, the absence of any correlation between the experimental amplitudes indicates that the spectra of the two materials are rather different. A comparative analysis of calculated atomic displacements allows to account for the low-energy IR modes which appear in the spectrum of CaMnO$_3$ as weak peaks. More generally, the present LDC shows that most of the IR modes differ significantly from PrMnO$_3$ to CaMnO$_3$, and correspond to rather complicated atomic vibrations that cannot be assigned to one type of vibration (external, bending, or stretching modes). So, as expected, it emerges that the two compounds have a different structure; nevertheless, CaMnO$_3$ does not seem to have a structure much closer to the ideal cubic perovskite structure than PrMnO$_3$

\begin{acknowledgments}
The authors are grateful to P. Thibaudeau and V. Ta Phuoc for fruitful discussions, C. Autret-Lambert and V. Briz\'{e} for having synthesized the samples, and M. Zaghrioui for Raman spectrocopy data. This work was supported in part by CAPES-COFECUB (project 500/01).
\end{acknowledgments}


\references
\bibitem{Dagotto03} For a review, see for example: E. Dagotto, \textit{Nanoscale Phase Separation and Colossal Magnetoresistance} (Springer-Verlag Berlin Heidelberg 2003).
\bibitem{Jirak85} Z. Jir\'{a}k, S. Krupi\v{c}ka, Z. \v{S}im\v{s}a, M. Dlouh\'{a}, and S. Vratislav, J. Magn. Magn. Mater. \textbf{53}, 153 (1985).
\bibitem{Yoshizawa95} H. Yoshizawa, H. Kawano, Y. Tomioka, and Y. Tokura, Phys. Rev. B \textbf{52}, R13 145 (1995).
\bibitem{Tomioka96} Y. Tomioka, A. Asamitsu, H. Kuwahara, Y. Moritomo, and Y. Tokura, Phys. Rev. B \textbf{53}, R1689 (1996).
\bibitem{Martin99} C. Martin, A. Maignan, M. Hervieu, and B. Raveau, Phys. Rev. B \textbf{60}, 12 191 (1999).
\bibitem{Zimmermann01} M. V. Zimmermann, C. S. Nelson, J. P. Hill, Doon Gibbs, M. Blume, D. Casa, B. Keimer, Y. Murakami, C. -C. Kao, C. Venkataraman, T. Gog, Y. Tomioka, and Y. Tokura, Phys. Rev. B \textbf{64}, 195133 (2001). 
\bibitem{Tomioka95} Y. Tomioka, A. Asamitsu,  Y. Moritomo, and Y. Tokura, J. Phys. Soc. Jpn. \textbf{64}, 3626 (1995).
\bibitem{Lees95} M. R. Lees, J. Barratt, G. Balakrishnan, D. McK. Paul, and M. Yethiraj, Phys. Rev. B \textbf{52}, R14 303 (1995).
\bibitem{Tokunaga98} M. Tokunaga, N. Miura, Y. Tomioka, and Y. Tokura, Phys. Rev. B \textbf{57}, 5259 (1998).
\bibitem{Anane99} A. Anane, J.-P. Renard, L. Reversat, C. Dupas, P. Veillet, M. Viret, L. Pinsard, and A. Revcolevschi, Phys. Rev. B \textbf{59}, 77 (1999).
\bibitem{Asamitsu97} A. Asamitsu, Y. Tomioka, H. Kuwahara, and Y. Tokura, Nature \textbf{388}, 50 (1997).
\bibitem{Ponnambalam99} V. Ponnambalam, Sachin Parashar, A. R. Raju, and C. N. R. Rao, Appl. Phys. Lett. \textbf{74}, 206 (1999).
\bibitem{Parashar00} S. Parashar, E. E. Ebenso, A. R. Raju, and C. N. R. Rao, Sol. Stat. Comm. \textbf{114}, 295 (2000).
\bibitem{Stankiewicz00} Jolanta Stankiewicz, Javier Ses\'{e}, Joaqu\'{i}n Garc\'{i}a, Javier Blasco, and Conrado Rillo, Phys. Rev. B \textbf{61}, 11 236 (2000).
\bibitem{Parashar04} Sachin Parashar, L. Sudheendra, A. R. Raju, and C. N. R. Rao, J. Appl. Phys. \textbf{95}, 2181 (2004).
\bibitem{Westhauser06} W. Westh\"{a}user, S. Schramm, J. Hoffmann, and C. Jooss, Eur. Phys. J. B \textbf{53}, 323 (2006).
\bibitem{Hwang95} H. Y. Hwang, T. T. M. Palstra, S-W. Cheong, and B. Batlogg, Phys. Rev. B \textbf{52}, 15 046 (1995).
\bibitem{Moritomo97} Y. Moritomo, H. Kuwahara, Y. Tomioka, and Y. Tokura, Phys. Rev. B \textbf{55}, 7549 (1997).
\bibitem{Kiryukhin97} V. Kiryukhin, D. Casa, J. P. Hill, B. Kelmer, A. Vigliante, Y. Tomioka, and Y. Tokura, Nature \textbf{386}, 813 (1997).
\bibitem{Ogawa98} K. Ogawa, W. Wei, K. Miyano, Y. Tomioka, and Y. Tokura, Phys. Rev. B \textbf{57}, R15 033 (1998).
\bibitem{Okimoto99} Y. Okimoto, Y. Tomioka, Y. Onose, Y. Otsuka, and Y. Tokura, Phys. Rev. B \textbf{59}, 7401 (1999).
\bibitem{TaPhuoc03} V. Ta Phuoc, R. Sopracase, G. Gruener, J. C. Soret, F. Gervais, A. Maignan, and C. Martin, Mater. Sci. Eng. B \textbf{104}, 131 (2003).
\bibitem{Cherepanov} V. A. Cherepanov, L. Yu. Barkhatova, A. N. Petrov, and V.I. Voronin, J. Solid State Chem. \textbf{118}, 53 (1995). 
\bibitem{Jirak97a} Z. Jir\'{a}k, J. Hejtm\'{a}nek, K. Kn\'{i}\v{z}ek, and R. Sonntag, J. Solid State Chem. \textbf{132}, 98 (1997).
\bibitem{Alonso00} J. A. Alonso, M. J. Mart\'{i}nez-Lope, M. T. Casais, and M. T. Fern\'{a}ndez-D\'{i}az, Inorg. Chem. \textbf{39}, 917 (2000).
\bibitem{Martin01} L. Mart\'{i}n-Carr\'{o}n, and A. de Andr\'{e}s, J. Alloys Comp. \textbf{323-324}, 417 (2001).
\bibitem{Jirak97b} Z. Jir\'{a}k, J. Hejtm\'{a}nek, E. Pollert, M. Mary\v{s}ko, M. Dlouh\'{a}, and S. Vratislav, J. Appl. Phys. \textbf{81}, 5790 (1997).
\bibitem{Zhou06} Qingdi Zhou, and Brendan J. Kennedy, J. Phys. Chem. Solids \textbf{67}, 1595 (2006).
\bibitem{Blasco00} J. Blasco, C. Ritter, J. Garc\'{i}a, J. M. de Teresa, J. P\'{e}rez-Cacho, and M. R. Ibarra, Phys. Rev. B \textbf{62}, 5609 (2000).
\bibitem{Jorge} M. E. Melo Jorge, A. Correia dos Santos, and M. R. Nunes, Int. J. Inorg. Mater. \textbf{3}, 915 (2001).
\bibitem{Iliev06a} M. N. Iliev, M. V. Abrashev, J. Laverdi\`{e}re, S. Jandl, M. M. Gospodinov, Y.-Q. Wang, and Y.-Y. Sun, Phys. Rev. B \textbf{73}, 064302 (2006).
\bibitem{Laverdiere06a} J. Laverdi\`{e}re, S. Jandl, A. A. Mukhin, V. Yu. Ivanov, V. G. Ivanov, and M. N. Iliev, Phys. Rev. B \textbf{73}, 214301 (2006).
\bibitem{Laverdiere06b} J. Laverdi\`{e}re, S. Jandl, A. A. Mukhin, and V. Yu. Ivanov, Eur. Phys. J. B \textbf{54}, 67 (2006). 
\bibitem{Martin03} L. Mart\'{i}n-Carr\'{o}n, J. S\'{a}nchez-Ben\'{i}tez, and A. de Andr\'{e}s, J. Solid State Chem. \textbf{171}, 313 (2003).
\bibitem{Martin02} L. Mart\'{i}n-Carr\'{o}n, A. de Andr\'{e}s, M. J. Mart\'{i}nez-Lope, M. T. Casais, and J. A. Alonso, Phys. Rev. B \textbf{66}, 174303 (2002). 
\bibitem{Liarokapis99} E. Liarokapis, Th. Leventouri, D. Lampakis, D. Palles, J. J. Neumeier, and D. H. Goodwin, Phys. Rev. B \textbf{60}, 12 758 (1999).
\bibitem{Granado01} E. Granado, N. O. Moreno, H. Martinho, A. Garc\'{i}a, J. A. Sanjurjo, I. Torriani, C. Rettori, J. J. Neumeier, and S. B. Oseroff, Phys. Rev. Lett. \textbf{86}, 5385 (2001).
\bibitem{Abrashev02} M. V. Abrashev, J. B\"{a}ckstr\"{o}m, L. B\"{o}rjesson, V. N. Popov, R. A. Chakalov, N. Kolev, R.-L. Meng, and M. N. Iliev, Phys. Rev. B \textbf{65}, 184301 (2002).
\bibitem{Iliev03} M. N. Iliev, M. V. Abrashev, V. N. Popov, and V. G. Hadjiev, Phys. Rev. B \textbf{67}, 212301 (2003).
\bibitem{Jung98} J. H. Jung, K. H. Kim, T. W. Noh, E. J. Choi, and Jaejun Yu, Phys. Rev. B \textbf{57}, R11 043 (1998).
\bibitem{Fedorov99} I. Fedorov, J. Lorenzana, P. Dore, G. De Marzi, P. Maselli, P. Calvani, S.-W. Cheong, S. Koval, and R. Migoni, Phys. Rev. B \textbf{60}, 11 875 (1999).
\bibitem{Douy01} Andr\'{e} Douy, Int. J. Inorg. Mat. \textbf{3}, 699 (2001).
\bibitem{Poeppelmeier82} K. R. Poeppelmeier, M. E. Leonowicz, and J. M. Longo, J. Solid State Chem. \textbf{44}, 89 (1982).
\bibitem{Gale97} J. D. Gale, J. Chem. Soc., Faraday Trans. \textbf{93}, 629 (1997).
\bibitem{Couzi72} Michel Couzi, and Pham Van Huong, Journal de Chimie Physique, 1339 (1972).
\bibitem{Atanassova94} Y. K. Atanassova, V. G. Hadjiev, P. Karen, and A. Kjekshus, Phys. Rev. B \textbf{50}, 586 (1994).
\bibitem{Popov95} V. N. Popov, J. Phys.: Condens. Matter \textbf{7}, 1625 (1995).
\bibitem{Lewis85} G. V. Lewis, and C. R. A. Catlow, J. Phys. C: solid State Phys. \textbf{18}, 1149 (1985).
\bibitem{Cherry95} M. Cherry, M. S. Islam, and C. R. A. Catlow, J. Solid State Chem. \textbf{118}, 125 (1995).
\bibitem{Gale96} J. D. Gale, philos. Mag. \textbf{73}, 3 (1996).
\bibitem{Islam96} M. S. Islam, M. Cherry, and C. R. A. Catlow, J. Solid State Chem. \textbf{124}, 230 (1996).
\bibitem{Iliev98} M. N. Iliev, M. V. Abrashev, H.-G. Lee, V. N. Popov, Y. Y. Sun, C. Thomsen, R. L. Meng , and C. W. Chu, Phys. Rev. B \textbf{57}, 2872 (1998).
\bibitem{Smirnova99} I.S. Smirnova, Physica B \textbf{262}, 247 (1999).
\bibitem{Kroumova} E. Kroumova, M. I. Aroyo, J. M. Perez-Mato, A. Kirov, C. Capillas, S. Ivantchev, and H. Wondratschek, Phase Transitions \textbf{76}, 155 (2003), internet address: http://www.cryst.ehu.es
\bibitem{Chen92} H. Chen and J. Callaway, Phys. Rev. B \textbf{45}, 2085 (1992).
\bibitem{Koval95} S. Koval and R. Migoni, Phys. Rev. B \textbf{51}, 6634 (1995).
\bibitem{Born} M. Born and K. Huang, \textit{Dynamical Theory of Cristal Lattices} (Oxford University Press, 2002), p. 336.
\end{document}